\documentclass[twocolumn,letter,a4paper]{jpsj3}
\usepackage{amsmath,amssymb,bm}
\usepackage{txfonts}
\usepackage[usenames]{color}

\def\journal #1#2#3#4{#1 {\bf #2} (#4) #3}

\title{Electron Correlation Induced Spontaneous Symmetry Breaking\\ and Weyl Semimetal Phase in a Strongly Spin-Orbit Coupled System}

\author{\name{Akihiko \surname{Sekine}}\thanks{E-mail: sekine@imr.tohoku.ac.jp} and \name{Kentaro \surname{Nomura}}}
\inst{Institute for Materials Research, Tohoku University, Sendai 980-8577, Japan} 

\abst{We study theoretically the electron correlation effect in a three-dimensional (3D) Dirac fermion system which describes a topologically nontrivial state.
It is shown within the mean-field approximation that time-reversal and inversion symmetries of the system are spontaneously broken in the region where both spin-orbit coupling and electron correlation are strong.
This phase is considered as an analog of that in the lattice quantum chromodynamics (QCD).
It is also shown that in the presence of magnetic impurities, electron correlation enhances the appearance of the Weyl semimetal phase between the topological insulator phase and the normal insulator phase.}

\kword{topological insulator, Weyl semimetal, spontaneous symmetry breaking, electron correlation, Wilson fermion}

\begin{document}
\maketitle
Recent discovery of topologically nontrivial phases has moved modern physics to a new direction\cite{Hasan2010,Qi2011}.
Especially, the experimental realization of 3D topological insulators\cite{Hsieh2008,Zhang2009} has accelerated the researches in this field.
It is known that strong spin-orbit coupling is essential to realize a topologically nontrivial state.
In the presence of inversion symmetry, this nontrivial state is characterized by the change in the parity of the lowest unoccupied band from even to odd due to strong spin-orbit coupling.
On the other hand, strong electron correlation has been a central issue in condensed matter physics.
Recent years, intensive studies have been done in systems where both strong spin-orbit coupling and strong electron correlation exist, triggered by the discovery of a novel Mott-insulating state in iridates\cite{Kim2008,Jackeli2009,Shitade2009,Watanabe2010}.
The motivation of these studies are in the searches for novel phases due to the interplay between spin-orbit coupling and electron correlation.
Preceding studies of the electron correlation effect on topological insulators have mainly focused on the competition between the spin (or charge) ordered phase and the topological insulator phase of Hubbard-like models on honeycomb lattices\cite{Raghu2008,Meng2010,Rachel2010,Varney2010,Hohenadler2011,Yamaji2011,Zheng2011,Yu2011}, other 2D lattices\cite{Sun2009,Wen2010,Yoshida2012} and 3D lattices\cite{Zhang2009a,Pesin2010,Mong2010,Kurita2011}.

Another topological phase, the Weyl semimetal, where gapless linear dispersions exist in a 3D system, has gathered attentions these days.
Near the band-touching points (Weyl points), the excitations are described by the massless Dirac-Weyl Hamiltonian.
This quasiparticle, the Weyl fermion, is assigned a chirality, and the bulk band gap opens only if the two Weyl fermions with opposite chirality meet each other.
This topological behavior originates in the nonzero Berry curvature enclosing a Weyl point\cite{Wan2011}.
The realization of the Weyl semimetal phase requires breaking of either time-reversal or inversion symmetry\cite{Murakami2007,Burkov2011,Burkov2011a,Delplace2012,Halasz2012}.
It is remarkable that the Weyl semimetal phase is predicted in strongly correlated pyrochlore iridates\cite{Wan2011}.
This suggests that topological phases emerge in strongly correlated $d$-electron systems.

In this Letter, we focus on the electron correlation effect in a 3D Dirac fermion system which describes a topologically nontrivial state, and analyze it within the mean-field approximation.
We assume the strength of the Coulomb interaction between the bulk electrons as a constant for simplicity, although it would be correct to adopt $1/r$ Coulomb interaction because the screening effect is considered to be weak in Dirac fermion systems.
Further we consider the case that the magnetic impurities are doped into the system, and discuss the emergence of the Weyl semimetal phase.

Let us consider a 3D model on a cubic lattice.
We start from the 3D Wilson fermion, which Hamiltonian is given by
\begin{equation}
\begin{aligned}
\mathcal{H}_{0}(\bm{k})=\sum_{\mu=1}^4R_\mu(\bm{k})\cdot\alpha_\mu\label{H_0}
\end{aligned}
\end{equation}
with
\begin{equation}
\left\{
\begin{aligned}
R_j(\bm{k})&=\sin k_j,\\
R_4(\bm{k})&=m_0+r{\sum}_{j}\left(1-\cos k_j\right),
\end{aligned}
\right.
\end{equation}
where $j\ (=1,2,3)$ denotes spacial axis, $r>0$, and $\alpha_\mu$ are the standard (Dirac) gamma matrices,
\begin{equation}
\begin{aligned}
\alpha_j=
\begin{bmatrix}
0 & \sigma_j\\
\sigma_j & 0
\end{bmatrix},\ \ \ 
\alpha_4=
\begin{bmatrix}
1 & 0\\
0 & -1
\end{bmatrix},\ \ \ 
\alpha_5=
\begin{bmatrix}
0 & -i\\
i & 0
\end{bmatrix}.
\end{aligned}
\end{equation}
Here $\sigma_j$ are the Pauli matrices.
These matrices satisfy the anticommutation relation $\{\alpha_\mu,\alpha_\nu\}=2\delta_{\mu\nu}$.
The energy of this system is measured in unit of $v_{\mathrm{F}}/a$ with $v_{\mathrm{F}}$ and $a$ being the Fermi velocity and the lattice constant, respectively.
The Wilson fermion was originally proposed as a model which describes lattice fermions without doublers.
Although 2$^3=8$ fermion doublers are generated due to the priodicity of $\sin k_j$, the Wilson term $r\sum_{j}(1-\cos k_j)$ eliminates them.
Thus when $m_0=0$, massless Dirac fermion appears only at the point $\bm{k}=(0,0,0)$.
On the other hand, the Hamiltonian (\ref{H_0}) with $m_0<0$ can be regarded as a simplified one which describes 3D topological insulators such as Bi$_2$Se$_3$\cite{Zhang2009}.
In Eq. (\ref{H_0}), the spinor is written in the basis of $\left[c^\dag_{\bm{k}A\uparrow},c^\dag_{\bm{k}A\downarrow},c^\dag_{\bm{k}B\uparrow},c^\dag_{\bm{k}B\downarrow}\right]$, where $c^\dag$ is the creation operator of an electron, $A$, $B$ denote two orbitals, and $\uparrow$ ($\downarrow$) denotes up- (down-) spin\cite{Zhang2009}.

In the presence of time-reversal symmetry and inversion symmetry, the $Z_2$ invariant of the system is given by\cite{Fu2007,Fu2007a}
\begin{equation}
\begin{aligned}
(-1)^\nu=\prod_{i=1}^8\left\{-\mathrm{sgn}\left[R_4\left(\bm{\Lambda}_i\right)\right]\right\},
\end{aligned}
\end{equation}
where $\bm{\Lambda}_i$ are eight time-reversal invariant momenta. 
It is easily shown that if $0>m_0>-2r$ or $-4r>m_0>-6r$ ($m_0>0$, $-2r>m_0>-4r$, or $-6r>m_0$), the system is topologically nontrivial (trivial).

Next we consider the Coulomb interaction between the bulk electrons, which is given by
\begin{equation}
\begin{aligned}
\mathcal{H}_{\mathrm{int}}=\frac{1}{2N}\sum_{\bm{k},\bm{k}',\bm{q}}\sum_{\alpha,\beta}V(\bm{q})c^\dag_{\bm{k}+\bm{q}\alpha}c^\dag_{\bm{k}'-\bm{q}\beta}c_{\bm{k}'\beta}c_{\bm{k}\alpha},\label{Coulomb}
\end{aligned}
\end{equation}
where $\alpha$ and $\beta$ denote the component of spinor, $(A,\uparrow)$, $(A,\downarrow)$, $(B,\uparrow)$, $(B,\downarrow)$. 
Although $V(\bm{q})=4\pi/q^2$ in the original form, we assume that $V(\bm{q})=V=\mathrm{const.}$ for simplicity. 
In the mean-field approximation, only the Fock term gives the correction to the mass term as follows:
\begin{equation}
\begin{aligned}
\mathcal{H}^{\mathrm{MF}}_{\mathrm{int}}&=\frac{V}{2N}\sum_{\bm{k},\bm{k}'}\sum_{\alpha,\beta}\left[\left\langle c^\dag_{\bm{k}\alpha}c_{\bm{k}\beta}\right\rangle \left\langle c^\dag_{\bm{k}'\beta}c_{\bm{k}'\alpha}\right\rangle -2\left\langle c^\dag_{\bm{k}\alpha}c_{\bm{k}\beta}\right\rangle c^\dag_{\bm{k}'\beta}c_{\bm{k}'\alpha}\right],\label{int_MF}
\end{aligned}
\end{equation}
where we have retained only the terms which satisfy $\bm{q}=\bm{k}'-\bm{k}$. 
Here we assume that $\hat{\Delta}_{\alpha\beta}\equiv \frac{1}{N}\sum_{\bm{k}}\left\langle c^\dag_{\bm{k}\alpha}c_{\bm{k}\beta}\right\rangle=(\Delta_4\alpha_4+\Delta_5\alpha_5)_{\alpha\beta}$. 
This assumption is based on that in QCD, $\left\langle \bar{\psi}_{\alpha}\psi_{\beta}\right\rangle=(\sigma I+i\pi\gamma_5)_{\alpha\beta}$ with $\sigma$ and $\pi$ being the chiral condensate and the pion condensate, respectively. 
Then Eq. (\ref{int_MF}) becomes
\begin{equation}
\begin{aligned}
\mathcal{H}^{\mathrm{MF}}_{\mathrm{int}}&=2NV\left(\Delta_4^2+\Delta_5^2\right)-V\sum_{\bm{k}}c^\dag_{\bm{k}}\left(\Delta_4\alpha_4+\Delta_5\alpha_5\right)c_{\bm{k}},\label{int_MF2}
\end{aligned}
\end{equation}
where we have defined $c^\dag_{\bm{k}}=\left[c^\dag_{\bm{k}A\uparrow},c^\dag_{\bm{k}A\downarrow},c^\dag_{\bm{k}B\uparrow},c^\dag_{\bm{k}B\downarrow}\right]$.

Combining Eqs. (\ref{H_0}) and (\ref{int_MF2}), we obtain the effective Hamiltonian
\begin{equation}
\begin{aligned}
\mathcal{H}_{\mathrm{eff}}(\bm{k})=\sum_{\mu=1}^5R_\mu(\bm{k})\cdot\alpha_\mu,\label{H_MF}
\end{aligned}
\end{equation}
where $R_j(\bm{k})=\sin k_j$, $R_4(\bm{k})=m_0-V\Delta_4+r{\sum}_{j}\left(1-\cos k_j\right)$, and $R_5(\bm{k})=-V\Delta_5$.
The energy eigenvalues of the effective Hamiltonian (\ref{H_MF}) is given by $\pm E_{\bm{k}}=\pm\sqrt{\sum_{\mu=1}^5\left[R_\mu(\bm{k})\right]^2}$.
Let us express the eigenfunctions of Eq. (\ref{H_MF}) as $\left|u^\pm_{\bm{k}\lambda}\right\rangle$ where $\pm$ denotes the positive or negative energy eigenvalue and $\lambda\ (=1,2)$ denotes the degeneracy of the energy eigenvalue.
Then $c_{\bm{k}}$ and $c^\dag_{\bm{k}}$ can be written as
\begin{equation}
\left\{
\begin{aligned}
c_{\bm{k}}&=\sum_\lambda\left[a_{\bm{k}\lambda}\left|u^+_{\bm{k}\lambda}\right\rangle+b_{\bm{k}\lambda}\left|u^-_{\bm{k}\lambda}\right\rangle\right],\\
c^\dag_{\bm{k}}&=\sum_\lambda\left[a^\dag_{\bm{k}\lambda}\left\langle u^+_{\bm{k}\lambda}\right|+b^\dag_{\bm{k}\lambda}\left\langle u^-_{\bm{k}\lambda}\right|\right],
\end{aligned}
\right.
\end{equation}
where $a^\dag_{\bm{k}\lambda}$ $\left(b^\dag_{\bm{k}\lambda}\right)$ is the creation operator of an electron in the positive (negative) energy band. Then the mean-field Hamiltonian of the system is given by
\begin{equation}
\begin{aligned}
H&=2NV\left(\Delta_4^2+\Delta_5^2\right)+\sum_{\bm{k}}c^\dag_{\bm{k}}\mathcal{H}_{\mathrm{eff}}(\bm{k})c_{\bm{k}}\\
&=2NV\left(\Delta_4^2+\Delta_5^2\right)+\sum_{\bm{k},\lambda}\left[E_{\bm{k}}a^\dag_{\bm{k}\lambda}a_{\bm{k}\lambda}-E_{\bm{k}}b^\dag_{\bm{k}\lambda}b_{\bm{k}\lambda}\right].
\end{aligned}
\end{equation}

In this paper, we consider the case of zero temperature and set the Fermi energy $\epsilon_{\mathrm{F}}=0$.
In this case, $\left\langle a^\dag_{\bm{k}\lambda}a_{\bm{k}'\lambda'}\right\rangle=0$ and $\left\langle b^\dag_{\bm{k}\lambda}b_{\bm{k}'\lambda'}\right\rangle=\delta_{\bm{k},\bm{k}'}\delta_{\lambda,\lambda'}$ are satified.
Thus it follows that
\begin{equation}
\begin{aligned}
\hat{\Delta}&=\frac{1}{N}\sum_{\bm{k},\lambda}\left|u^-_{\bm{k}\lambda}\right\rangle\left\langle u^-_{\bm{k}\lambda}\right|\\
&=\frac{1}{N}\sum_{\bm{k}}\frac{1}{2E_{\bm{k}}}\left[E_{\bm{k}}-\mathcal{H}(\bm{k})\right].
\end{aligned}
\end{equation}
The values of $\Delta_4$ and $\Delta_5$ are obtained by solving the following self-consistent equations:
\begin{equation}
\left\{
\begin{aligned}
\Delta_4&=\frac{1}{4}\mathrm{tr}\left(\alpha_4\hat{\Delta}\right)=-\frac{1}{N}\sum_{\bm{k}}\frac{R_4(\bm{k})}{2E_{\bm{k}}},\\
\Delta_5&=\frac{1}{4}\mathrm{tr}\left(\alpha_5\hat{\Delta}\right)=-\frac{1}{N}\sum_{\bm{k}}\frac{R_5(\bm{k})}{2E_{\bm{k}}}.
\end{aligned}\label{Delta_4-Delta_5}
\right.
\end{equation}
Solving these two equations is equivalent to obtaining the stationary point of the free energy per unit volume of the system $F(\Delta_4,\Delta_5)$:
\begin{equation}
\begin{aligned}
\frac{\partial F(\Delta_4,\Delta_5)}{\partial \Delta_4}=\frac{\partial F(\Delta_4,\Delta_5)}{\partial \Delta_5}=0,
\end{aligned}
\end{equation}
where
\begin{equation}
\begin{aligned}
F(\Delta_4,\Delta_5)=2V\left(\Delta_4^2+\Delta_5^2\right)-\frac{2}{N}\sum_{\bm{k}}E_{\bm{k}}.
\end{aligned}
\end{equation}

\begin{figure}[!t]
\begin{center}
\includegraphics[width=\columnwidth]{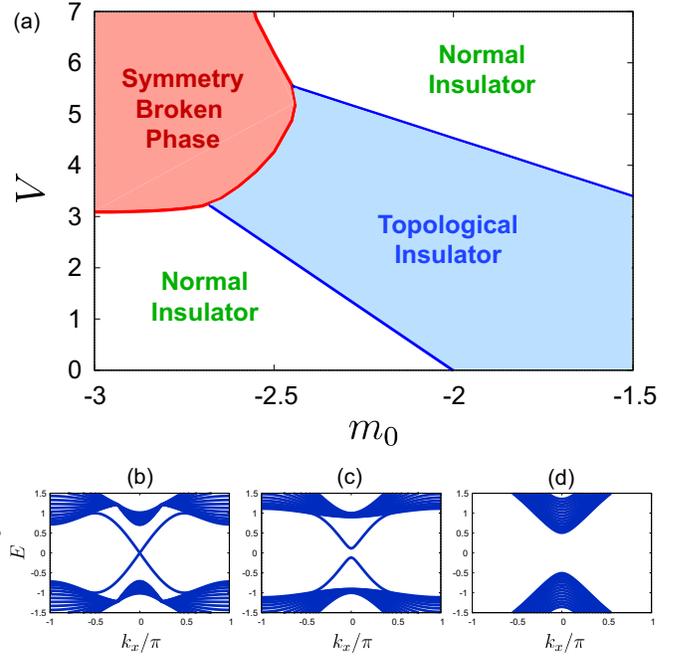}
\caption{(Color online) (a) Phase diagram.
The phase boundary between the topological insulator phase and the normal insulator phase is determined by the condition $m_{\rm eff}=0$ or $m_{\rm eff}=-2r$.
The phase boundary between the $\Delta_5\neq 0$ phase and the other phase is determined by the condition $\Delta_5=0$.
The $k_x$ dependence of surface spectra with $k_y=0$ at $(m_0,V)=(-2.5,3.7)$, $(-2.5,4.3)$ and $(-2.5,5.5)$ are shown in (b), (c) and (d), respectively. }
\end{center}
\vspace{-0.5cm}
\end{figure}

In the effective Hamiltonian, we define the effective mass:
\begin{equation}
\begin{aligned}
m_{\mathrm{eff}}=m_0-V\Delta_4.
\end{aligned}
\end{equation}
The phase diagram is shown in Fig. 1(a).
Throughout this paper, we set $r=1$.
Although this value is far from that which was obtained by an {\it ab initio} calculation\cite{Zhang2009} for Bi$_2$Se$_3$, we have confirmed that the value of $r$ does not change the results largely.
It was found that $\Delta_4$ is always negative, and the effective mass is a monotonically increasing function with respect to the electron correlation strength $V$.
As a result, there exists the critical strength $V_c$ where the effective mass changes its sign in the $m_0<0$ region.
As mentioned above, if $0>m_{\mathrm{eff}}>-2$ ($m_{\mathrm{eff}}>0$ or $-2>m_{\mathrm{eff}}>-4$), the system is identified as a topological insulator (normal insulator).
The value of $m_0$ for Bi$_2$Se$_3$ is estimated as about $-0.3$.

We see that two values of $\Delta_5$, namely $\Delta_5=0$ or $\Delta_5=\pm c$ ($c$ is real), can exist as the solutions of Eq. (\ref{Delta_4-Delta_5}).
The phase where $\Delta_5\neq 0$ is realized if $F(\Delta_4,\Delta_5\neq 0)<F(\Delta_4,\Delta_5=0)$.
For example, on the $m_0=-2.5$ line, $\Delta_5$ starts from 0 to take nonzero value with increasing $V$, and reaches the maxinum absolute value $\pm0.098$. Then $\Delta_5$ decreases toward zero.
The surface spectra in the topological insulator phase and the $\Delta_5\neq 0$ phase are shown in Fig. 1(b) and Fig. 1(c), (d), respectively.
In the $\Delta_5\neq 0$ phase, the energy gap of surface modes opens due to nonzero $\Delta_5$ and thus the system is insulating.
From the symmetry point of view, the matrix $\alpha_5$ breaks both time-reversal ($\mathcal{T}$) symmetry and inversion ($\mathcal{I}$) symmetry, i.e., $\mathcal{T}\alpha_5\mathcal{T}^{-1}=-\alpha_5$ and $\mathcal{I}\alpha_5\mathcal{I}^{-1}=-\alpha_5$ are satisfied, where $\mathcal{T}=\bm{1}\otimes(-i\sigma_2)\mathcal{K}$ and $\mathcal{I}=\sigma_3\otimes\bm{1}$.
This means that the spontaneous symmetry breaking occurs due to electron correlation.
It is known that a similar phase (in which parity and flavor symmetry are spotaneously broken) exists in the lattice QCD with Wilson fermion\cite{Aoki1984,Aoki1986,Sharpe1998}.
In Fig. 1(a), the $\Delta_5\neq 0$ phase is realized in the region where both $|m_0|$ and $V$ are large.
Negative sign of $m_0$ originates in the level crossing of the two orbitals induced by spin-orbit coupling.
Therefore, the value of $|m_0|$ can be considered as the strength of spin-orbit coupling\cite{Zhang2009}.
It can be said that this novel phase may emerge in systems where both spin-orbit couling and electron correlation are strong.

\begin{figure}[!t]
\begin{center}
\includegraphics[width=\columnwidth,clip]{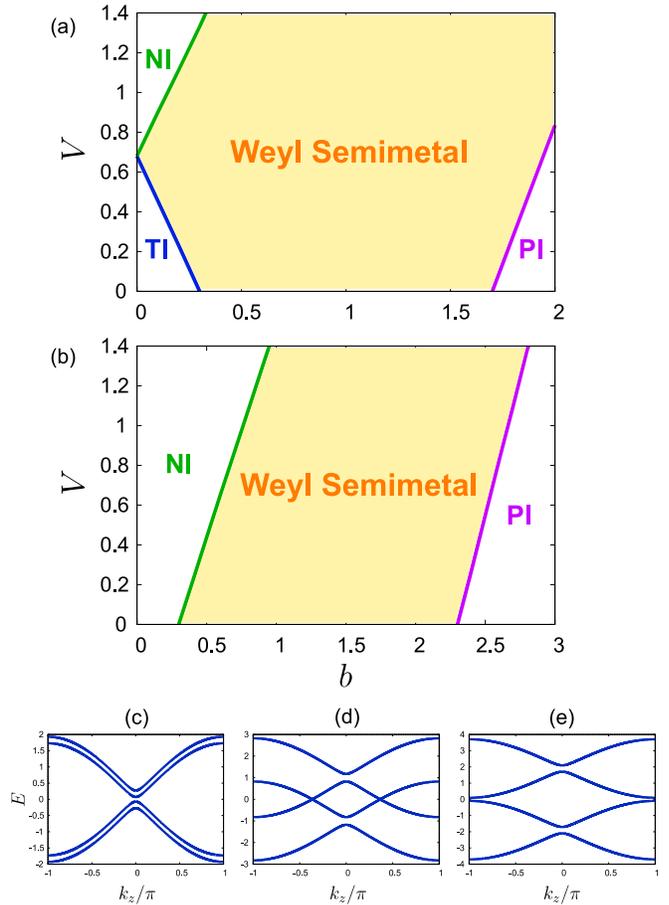}
\caption{(Color online) Phase diagram calculated with the $\Sigma_{12}$ term.
(a) In the case of $m_0=-0.30$, (b) $m_0=+0.30$.
 The topological insulator (TI) phase and the normal insulator (NI) phase are determined by the sign of the effective mass.
The Weyl semimetal (polarized insulator (PI)) phase is obtained as the phase in which the Weyl points exist (do not exsist).
The $k_z$ dependence of the energy bands (Eq. (\ref{energyband})) in the case of $m_0=-0.30$ with ($k_x$,$k_y$)=(0,0) at $(b,V)=(0.10,0.30)$, $(1.0,0.30)$ and $(1.90,0.30)$ are shown in (c), (d) and (e), respectively.}
\end{center}
\vspace{-0.5cm}
\end{figure}

Next we consider to add a time-reversal symmetry breaking term\cite{Burkov2011}.
This term can be regarded as an external magnetic field along the $z$-axis or magnetic impurities.
In this case, the effective Hamiltonian is written as
\begin{equation}
\begin{aligned}
\mathcal{H}_{\mathrm{eff}}(\bm{k})=\sum_{\mu=1}^5R_\mu(\bm{k})\cdot\alpha_\mu+b\Sigma_{12},\label{Heff2}
\end{aligned}
\end{equation}
where $\Sigma_{12}=-\frac{i}{2}\left[\alpha_1,\alpha_2\right]=\bm{1}\otimes\sigma_3$.
We can obtain the eigenvalues of this effective Hamiltonian analytically, which leads to
\begin{equation}
\begin{aligned}
E(\bm{k})=\pm\sqrt{\sum_{\mu=1}^2\left[R_\mu(\bm{k})\right]^2+\left(\sqrt{\sum_{\mu=3}^5\left[R_\mu(\bm{k})\right]^2}\pm b\right)^2}.\label{energyband}
\end{aligned}
\end{equation}
The Weyl points, at which the two bands touch, appear where the wave vector $\bm{k}$ satisfies the condition $b^2=\left[m_{\mathrm{eff}}+r{\sum}_{j}\left(1-\cos k_j\right)\right]^2+\left(V\Delta_5\right)^2+\sin^2 k_z$ and $\sin k_x=\sin k_y=0$.
$k_x$ and $k_y$ can take the value 0 or $\pi$.
Setting $|m_0|$ small, we concentrate on the Weyl points which appear on the $(k_x,k_y)=(0,0)$ line.
The values of $\Delta_4$ and $\Delta_5$ are obtained by solving two equations self-consistently: $\partial F(\Delta_4,\Delta_5)/\partial \Delta_4=\partial F(\Delta_4,\Delta_5)/\partial \Delta_5=0$.
Here $F$ is the free energy per unit volume at zero temperature, which is defined by
\begin{equation}
\begin{aligned}
F(\Delta_4,\Delta_5)=2V\left(\Delta_4^2+\Delta_5^2\right)+\frac{1}{N}\sum_{\bm{k}}\sum_{\lambda=1}^2E^\lambda_{\bm{k}},
\end{aligned}
\end{equation}
where $E^\lambda_{\bm{k}}$ are the engative two energy bands of Eq. (\ref{energyband}).

The phase diagram calculated with the $\Sigma_{12}$ term is shown in Fig. 2.
First we set $m_0=-0.30$, intending Bi$_2$Se$_3$. [see Fig. 2(a)].
As in the case of $b=0$, the effective mass $m_{\mathrm{eff}}$ is a monotonically increasing function of $V$.
We see that the Weyl semimetal phase emerges between the topological and normal insulator phases, as the perturbation term $b$ increases.
Although the Hamiltonian (\ref{Heff2}) does not have time-reversal symmetry, we call the region where the effective mass is negative the topological insulator phase.
The transition from the topological insulator phase to the other phase occurs when the band gap closes.
Thus as far as the band gap is open, we call the phase in which the effective mass is negative the topological insulator phase, regardless of the loss of time-reversal symmetry.

The Weyl semimetal phase is defined as a phase where the gapless dispersions exist on the $k_z$-axis.
In this phase, the effective mass changes its sign.
However, once the band gap closes, the sign of the effective mass has no importance.
At fixed values of $b$, the Weyl points first arise at $k_z=0$ with increasing $V$.
The Weyl points split and start to move toward a certain value of $k_z$, then go back to $k_z=0$ and the band gap opens.
At fixed values of $V$, the Weyl points first appear at $k_z=0$ with increasing $b$, then the Weyl points split and start to move toward $k_z=\pm\pi$.
These points disappear after reaching $k_z=\pm\pi$ and the band gap opens.
We call this state the polarized insulator, because this phase is realized with large $b$.
When $V=0$, we can obtain the value of the phase boundary analytically: $b=|m_0|=0.30$ ($k_z=0$) and $b=m_0+2r=1.70$ ($k_z=\pi$).
The four energy bands (Eq. (\ref{energyband})) at $(b,V)=(0.10,0.30)$, $(1.0,0.30)$ and $(1.90,0.30)$ on the phase diagram (Fig. 2(a)) are plotted as a function of $k_z$ in Fig. 2(c), (d) and (e), respectively.

Next we set $m_0=+0.30$ [see Fig. 2(b)].
In this case, the effective mass $m_{\mathrm{eff}}$ always takes positive value and $m_{\mathrm{eff}}$ is a monotonically increasing function of $V$.
Thus the Weyl semimetal phase begins to arise with finite $b$ due to the condition $b^2=\left[m_{\mathrm{eff}}+r{\sum}_{j}\left(1-\cos k_j\right)\right]^2+\sin^2 k_z$, in contrast to the case of $m_0=-0.30$.
The behavior of the Weyl points with respect to $b$ and $V$ is the same as the case of $m_0=-0.30$.
When $V=0$, we can obtain the value of the phase boundary analytically: $b=m_0=0.30$ ($k_z=0$) and $b=m_0+2r=2.30$ ($k_z=\pi$).
\begin{figure}[!t]
\begin{center}
\includegraphics[width=\columnwidth]{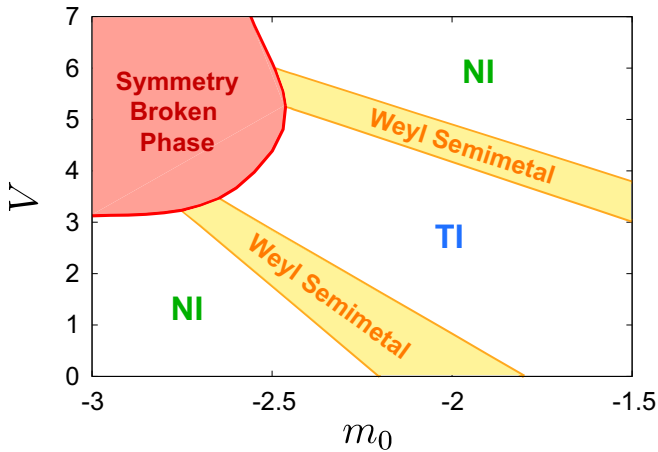}
\caption{(Color online) Phase diagram with $b=0.20$. In the presence of magnetic impurities, the regions where the absolute value of $m_{\rm eff}$ and $m_{\rm eff}+2r$ are small in Fig. 1(a) changes to the Weyl semimetal phase.}
\end{center}
\vspace{-0.5cm}
\end{figure}

Finally, the phase diagram with $b$ fixed to 0.20 is shown in Fig. 3.
When $b$ is small, there exists the Weyl semimetal phase only in the $m_0<0$ region.
This emergence of the Weyl semimetal phase results from the change in the sign of $m_{\mathrm{eff}}$ and $m_{\mathrm{eff}}+2r$.
We see that the $\Delta_5\neq 0$ phase is robust against the perturbation $b$. If $\Delta_5$ is not zero, the band gap opens and therefore the Weyl semimetal phase cannot exist.

In a realistic system, for example in Bi$_2$Se$_3$, the value of $m_0$ is estimated as about $-0.3$.
Our calculation shows that the critical strength $V_c$ at which the Weyl semimetal phase appears is about $0.7\hbar v_{\rm F}/a=0.7$ [eV], when $\hbar v_{\rm F}=3$ [eV$\cdot{\rm \AA}$] and $a=3$ [${\rm \AA}$] are assumed.
This value is considered to be too large for $p$-electron systems.
Therefore, a possible way to realize the Weyl semimetal phase in 3D topological insulators such as Bi$_2$Se$_3$ is to make the bulk band gap ($\simeq |m_0|$) smaller without the big change in the strength of Coulomb interaction between the bulk electrons.
This experimental procedure has been realized in a solid-solution system BiTl(S$_{1-x}$Se$_x$)$_2$\cite{Xu2011,Sato2011}.

To summarize, we have studied the electron correlation effect in a 3D topological insulator which is described by the Wilson fermion within the mean-field approximation.
As the correlation strength increases, the topological insulator phase changes to the normal insulator phase and vice versa.
In the region where both spin-orbit coupling and electron correlation are strong, time-reversal and inversion symmetries of the system are spontaneously broken. This phase is considered as an analog of that in the lattice QCD.
By adding a time-reversal symmetry breaking term, the Weyl semimetal phase emerges between the topological and normal insulator phases.
In our calculation, the bulk band gap monotonically increases and becomes infinity in the limit of strong correlation.
The analysis of not only the bulk but also the surface states from the strong correlation limit is needed to understand the correation effect in 3D topological insulators.

\begin{acknowledgments}
This work was supported by Grant-in-Aid for Scientific Research (No. 24740211) from the Ministry of Education, Culture, Sports, Science and Technology (MEXT), Japan. A.S. is finacially supported by the global COE program of MEXT, Japan.
\end{acknowledgments}

\end{document}